\documentclass[11pt,a4paper]{amsart}
\usepackage[margin=25mm]{geometry}
\usepackage[numbers]{natbib}
\usepackage{hyperref}

\usepackage{amssymb,amsmath}
\usepackage{amsthm}
\usepackage{bbm, bm}
\usepackage{stmaryrd}
\usepackage[usenames,dvipsnames]{xcolor}
\usepackage{color}
\usepackage{graphicx}
\usepackage{caption}
\usepackage{float}
\usepackage{wrapfig}
\usepackage{enumerate,enumitem}  

\usepackage{lineno}

\usepackage{abstract}

\input xy
\xyoption{all} 

\numberwithin{equation}{section}

\theoremstyle{definition}

\newcommand{\rmd}{\textnormal{d}}

\DeclareMathOperator{\Vect}{Vect}

\DeclareMathOperator{\Span}{Span}

\font\black=cmbx10 \font\sblack=cmbx7 \font\ssblack=cmbx5 \font\blackital=cmmib10  \skewchar\blackital='177
\font\sblackital=cmmib7 \skewchar\sblackital='177 \font\ssblackital=cmmib5 \skewchar\ssblackital='177
\font\sanss=cmss10 \font\ssanss=cmss8 
\font\sssanss=cmss8 scaled 600 \font\blackboard=msbm10 \font\sblackboard=msbm7 \font\ssblackboard=msbm5
\font\caligr=eusm10 \font\scaligr=eusm7 \font\sscaligr=eusm5  \font\fraktur=eufm10
\font\sfraktur=eufm7 \font\ssfraktur=eufm5 
\font\bsymb=cmsy10 scaled\magstep2
\def\all#1{\setbox0=\hbox{\lower1.5pt\hbox{\bsymb
       \char"38}}\setbox1=\hbox{$_{#1}$} \box0\lower2pt\box1\;}
\def\exi#1{\setbox0=\hbox{\lower1.5pt\hbox{\bsymb \char"39}}
       \setbox1=\hbox{$_{#1}$} \box0\lower2pt\box1\;}

\def\tx#1{{\fam0\relax#1}}

\newfam\bifam
\textfont\bifam=\blackital \scriptfont\bifam=\sblackital \scriptscriptfont\bifam=\ssblackital

\newfam\blfam
\textfont\blfam=\black \scriptfont\blfam=\sblack \scriptscriptfont\blfam=\ssblack

\newfam\bbfam
\textfont\bbfam=\blackboard \scriptfont\bbfam=\sblackboard \scriptscriptfont\bbfam=\ssblackboard

\newfam\ssfam
\textfont\ssfam=\sanss \scriptfont\ssfam=\ssanss \scriptscriptfont\ssfam=\sssanss
\def\sss#1{{\fam\ssfam\relax#1}}

\newfam\clfam
\textfont\clfam=\caligr \scriptfont\clfam=\scaligr \scriptscriptfont\clfam=\sscaligr

\newfam\frfam
\textfont\frfam=\fraktur \scriptfont\frfam=\sfraktur \scriptscriptfont\frfam=\ssfraktur

\def\hpb#1{\setbox0=\hbox{${#1}$}
    \copy0 \kern-\wd0 \kern.2pt \box0}
\def\vpb#1{\setbox0=\hbox{${#1}$}
    \copy0 \kern-\wd0 \raise.08pt \box0}

\def\pmb#1{\setbox0\hbox{${#1}$} \copy0 \kern-\wd0 \kern.2pt \box0}
\def\pmbb#1{\setbox0\hbox{${#1}$} \copy0 \kern-\wd0
      \kern.2pt \copy0 \kern-\wd0 \kern.2pt \box0}
\def\pmbbb#1{\setbox0\hbox{${#1}$} \copy0 \kern-\wd0
      \kern.2pt \copy0 \kern-\wd0 \kern.2pt
    \copy0 \kern-\wd0 \kern.2pt \box0}
\def\pmxb#1{\setbox0\hbox{${#1}$} \copy0 \kern-\wd0
      \kern.2pt \copy0 \kern-\wd0 \kern.2pt
      \copy0 \kern-\wd0 \kern.2pt \copy0 \kern-\wd0 \kern.2pt \box0}
\def\pmxbb#1{\setbox0\hbox{${#1}$} \copy0 \kern-\wd0 \kern.2pt
      \copy0 \kern-\wd0 \kern.2pt
      \copy0 \kern-\wd0 \kern.2pt \copy0 \kern-\wd0 \kern.2pt
      \copy0 \kern-\wd0 \kern.2pt \box0}

\mathchardef\za="710B  
\mathchardef\zb="710C  
\mathchardef\zg="710D  
\mathchardef\zd="710E  
\mathchardef\zve="710F 
\mathchardef\zz="7110  
\mathchardef\zh="7111  
\mathchardef\zvy="7112 
\mathchardef\zi="7113  
\mathchardef\zk="7114  
\mathchardef\zl="7115  
\mathchardef\zm="7116  
\mathchardef\zn="7117  
\mathchardef\zx="7118  
\mathchardef\zp="7119  
\mathchardef\zr="711A  
\mathchardef\zs="711B  
\mathchardef\zt="711C  
\mathchardef\zu="711D  
\mathchardef\zvf="711E 
\mathchardef\zq="711F  
\mathchardef\zc="7120  
\mathchardef\zw="7121  
\mathchardef\ze="7122  
\mathchardef\zy="7123  
\mathchardef\zf="7124  
\mathchardef\zvr="7125 
\mathchardef\zvs="7126 
\mathchardef\zf="7127  
\mathchardef\zG="7000  
\mathchardef\zD="7001  
\mathchardef\zY="7002  
\mathchardef\zL="7003  
\mathchardef\zX="7004  
\mathchardef\zP="7005  
\mathchardef\zS="7006  
\mathchardef\zU="7007  
\mathchardef\zF="7008  
\mathchardef\zW="700A  
\mathchardef\zC="7009  

\newcommand{\be}{\begin{equation}}
\newcommand{\ee}{\end{equation}}

\newcommand{\bea}{\begin{eqnarray}}
\newcommand{\eea}{\end{eqnarray}}
\def\*{{\textstyle *}}
\newcommand{\R}{{\mathbb R}}

\newcommand{\s}{{\textstyle *}}

\def\Sec{\sss{Sec}}
\def\Vect{\sss{Vect}}

\def\sV{{\sss V}}

\def\xi{\tx{i}}

\def\s*{{\scriptstyle *}}

\newcommand{\beas}{\begin{eqnarray*}}
\newcommand{\eeas}{\end{eqnarray*}}

\title{Carrollian $\R^\times$-bundles II: Sigma Models on Event Horizons} 
\author{Andrew James Bruce }  
   \email{andrewjamesbruce@googlemail.com}

   \date{\today}
 
\begin{document}
 \maketitle
\vspace{-20pt}
\begin{abstract}{\noindent Carrollian field theories are usually understood as limits of relativistic theories. In this note, we use Carrollian $\R^\times$-bundles equipped with a principal connection to construct Carrollian sigma models intrinsically. The resulting theories are neither ``electric'' nor ``magnetic'' in the usual sense. As a physically suggestive example, we derive a Carrollian wave equation governing the dynamics of a scalar field on the event horizon of a Schwarzschild black hole. }\\

\noindent {\Small \textbf{Keywords:} Carrollian Geometry;~Carrollian Field Theories}

\end{abstract}
%
%
\section{Introduction}
Duval et al.\ \cite{Duval:2014a,Duval:2014b, Duval:2014} introduced the intrinsic definition of a \emph{Carrollian manifold}  as a smooth manifold equipped with a degenerate metric whose kernel is spanned by a nowhere vanishing complete vector field. Earlier works on Carrollian geometry include Lévy-Leblond \cite{Lévy-Leblond:1965}, Sen Gupta \cite{SenGupta:1966}, and Henneaux \cite{Henneaux:1979}. Lévy-Leblond showed that in the ultra-relativistic limit $c\rightarrow 0$, massive particles move at infinite speed and yet remain stationary; particles are ``frozen'' in space. Natural examples of Carrollian manifolds include null hypersurfaces such as punctured future or past light cones in Minkowski spacetime and the event horizon of a Schwarzschild black hole.  \par 
There has been a renewed interest in building (quantum) field theories that are invariant under Carrollian diffeomorphisms, driven by the study of field theories in extreme limits, see \cite{Correa:2024,Cotler:2024,Sharma:2025}, for example.  To build a Carrollian field theory, one method is to consider a field theory on a Bargmann manifold and reduce it along one of the null directions. In this note, we leverage the power of  Carrollian $\R^\times$-bundles (see \cite{Bruce:2025} for details) to build a classical sigma model. The theory is \emph{intrinsic} in the sense that it is built directly from the geometry and is not a limit of some relativistic theory. Moreover, the Lagrangian includes derivatives with respect to both space and time, a feature that is not usually present in  Carrollian theories. Broadly, a Carrollian field theory is referred to as ``electric'' if the dynamics are dominated by time evolution in the ultra-relativistic limit. Dual to that, a Carrollian field theory is referred to as ``magnetic'' if the dynamics strongly depend on spatial gradients. The theories we build are generally neither  ``electric'' nor ``magnetic'' as they do not arise as a limit.  We must highlight the work of Ciambelli \cite{Ciambelli:2024}, who showed that the ``electric'' and ``magnetic'' Carroll scalar actions can be combined on a generic Carrollian manifold. However, the  ``mixed theory'' cannot come from the limit $c\rightarrow 0$ of a relativistic scalar field theory; the intrinsic point of view is required and leads to non-trivial dynamics. Ciambelli's observations sit comfortably with the work presented here. Nevertheless, we emphasize that the models we construct are distinct from Ciambelli's and do not include his as a special case. We also mention Ecker et al. \cite{Ecker:2024}, who have also studied intrinsic Carrollian theories that have non-trivial dynamics.
\par 
A principal $\R^\times$-bundle $P$ is said to be a \emph{Carrollian $\R^\times$-bundle} if it is equipped with a degenerate metric $g$ such that $\ker(g) := \left \{X \in \Vect(P)  ~|~  g(X,-)=0 \right \} =  \Sec(\sV P)$, where $\sV P$ is the vertical bundle. Note that the fundamental vector field of the principal action $\Delta_P$, which we will refer to as the \emph{Euler vector field}, provides a canonical basis of $\Sec(\sV P)$. We principally restrict our attention to Carrollian $\R^\times$-bundles  with a chosen $\R^\times$-connection $(P, g, \Phi)$, such that $M$ is $n$-dimensional, the degenerate metric has signature $(1,\cdots,1,0)$, and the Euler vector field is a Killing vector field. As we will employ local coordinates $(x^a, t)$ on $P$, the signature of the metric has been chosen to align with the ordering of local coordinates, i.e., the kernel of $g$ lies along the fibre direction. In these adapted coordinates, $\Delta_P = t \partial_t$.  Note that the admissible coordinate changes are ``linearised Carrollian differomorphisms'', i.e., $x^{a'} = x^{a'}(x)$ and $t' = \phi(x)t$.  As shown in \cite{Bruce:2025}, any Carrollian bundle, i.e., a Carrollian manifold in which the associated foliation is a fibre bundle with typical fibre $\R$, can be (non-canonically) linearised, and that there is a fibre-preserving diffeomorphism from the linearised bundle to the initial Carrollian bundle. Coupled with the one-to-one correspondence between line bundles and principal $\R^\times$-bundle there is no real loss of generality. In effect, we are picking adapted coordinates to make the coordinate changes linear in the fibre coordinates. \par 
The connection one-form associated with $\Phi$ we denote as $\theta$. Note that this is a globally defined real-valued one-form, and is locally of the form $\theta = t^{-1}\rmd t + \rmd x^a A_a$. As first presented in \cite{Bruce:2025}, $(P, g, \Phi)$ comes with the non-degenerate Lorentzian metric  $G := g - \theta \otimes \theta$.  The connection is seen as part of the underlying geometry and not a dynamical field that is required by local symmetries or sourced by matter - its role is an intrinsic part of the Carrollian geometry.  Written out (locally) in block form we have
\begin{align*}
G = \begin{pmatrix}
g_M - A   A^T & -t^{-1} A \\
-t^{-1}A^T  & - t^{-2}
\end{pmatrix}
\,,
& &
G^{-1} = \begin{pmatrix}
g_M^{-1}  & -t g_M^{-1} A \\
-t A^Tg_M^{-1}  & - t^2+ t^2 A^Tg_M^{-1}A
\end{pmatrix}  \, ,
\end{align*}
where $g_M$ is the non-degenerate metric on the base manifold $M$. It is important to note that the gauge field has been chosen so that $G$ is of signature $(1,1,1,-1)$ (as we have ordered the coordinates). This can always be done as the connection is not dynamical and is a fixed part of the geometry.   Note that $\det(G) = - t^{-2}\,\det(g_M) $, and thus $\sqrt{|G|} = \sqrt{|g_M|}\,  t^{-1} $. It is important to remark the metric on $M$ and the connection allow for a global invariant measure to be constructed, and this is vital to building actions. By framing Carrollian geometry in terms of principal bundles, we naturally and geometrically circumvent the issue of defining measures on Carrollian manifolds.  
%
%
\section{The Carrollian Sigma Model}
The action we construct is a non-linear sigma model, where the fields are smooth maps $\Psi : P \rightarrow N$, and $(N, k)$ is a (pseudo-)Riemannian manifold. The total space of $P$ is equipped with the metric $G = g - \theta\otimes \theta$. This means we work with a fixed principal connection on $P$.  Equipping $N$ with local coordinates $y^i$, we write $\Psi^*(y^i) = \Psi^i(x,t)$.  Locally, the action is 
\begin{align}
&S = \int \rmd^n x\,  \rmd t \, \sqrt{|G|} \left(- \frac{1}{2} G^{\mu \nu}\partial_\nu \Psi^i \partial_\mu \Psi^j\, k_{ji}(\Psi)\right )\\ \nonumber 
&= \int \rmd^n x\,  \rmd t \, \sqrt{|g_M|}\,  t^{-1} \left(- \frac{1}{2} g_M^{a b}\partial_b \Psi^i \partial_a \Psi^j\, +   \, g_M^{ab} A_b  \partial_a \Psi^i \Delta_P \Psi^j  
  + \frac{1}{2} \big(1- g_M^{ab} A_a A_b\big)\Delta_P \Psi^i \Delta_P \Psi^j  \right ) k_{ji}(\Psi)\,.
\end{align}
The action consists of a standard kinetic term in three dimensions, a non-minimal coupling of scalar fields with the gauge field, and a non-standard kinetic term involving the Euler vector field. \par  

The equation of motion is
\begin{align}\label{eqn:ELSig}
&g_M^{ab}\nabla_b \nabla_a \Psi^i  +(1 - g_M^{ab}A_bA_a)\nabla_{\Delta_P}\nabla_{\Delta_P}  \Psi^i \\ \nonumber 
& +\Gamma^i_{jk}\left(g_M^{ab}\partial_b \Psi ^k \partial_a \Psi^j - 2 g_M^{ab}A_b \partial_a\Psi^k \Delta_P \Psi^j + (1- g_M^{ab}A_a A_b)\Delta_P \Psi^k \Delta_P \Psi^j  \right)=0\,.
\end{align}
Some comments are in order:
\begin{enumerate}
\item by construction, the action - and hence the equations of motion - are invariant under coordinate changes $x^{a'} = x^{a'}(x)$ and $t' = \phi(x)\,t$, i.e., ``linearised Carrollian'' transformations;
\item the theory is not ultralocal or ``frozen'' as the Euler--Lagrange equations \eqref{eqn:ELSig} contain first and second-order derivatives in both $x$ and $t$; and
\item if $\Psi^i$ is homogeneous and of weight $l$, i.e., $\Delta_P \Psi =l \, \Psi$, then the Euler--Lagrange equation becomes
\begin{align}
&g_M^{ab}\nabla_b \nabla_a \Psi^i  + l^2\,(1 - g_M^{ab}A_bA_a) \Psi^i \\ \nonumber 
& +\Gamma^i_{jk}\left(g_M^{ab}\partial_b \Psi ^k \partial_a \Psi^j - 2 l\,g_M^{ab}A_b \partial_a\Psi^k  \Psi^j + l^2\,(1- g_M^{ab}A_a A_b) \Psi^k  \Psi^j  \right)=0\,.
\end{align}
Thus, the time derivatives are replaced by an algebraic condition via $l$. Moreover, if $l=0$, then the theory reduces to a standard sigma model with source $(M,g_M)$.
\end{enumerate}
Note that the Euler--Lagrange equation \eqref{eqn:ELSig} explicitly depends on $t$ via the Euler vector field $\Delta_P$. In particular, there is no dependence on $t^{-1}$ in the Euler--Lagrange equations. Thus, provided $\Psi$ and its derivatives up to second order are regular as $t\rightarrow 0$, i.e., do not blow up as they approach $t=0$, then the Euler--Lagrange equations and their solutions can be ``analytically extended'' to the line bundle $L$ associated with $P$. That is, we can consider the theory on $L$ and have a well-defined theory that includes $t=0$; even if the action cannot be defined on $L$, the dynamics are perfectly well-defined.\par
The previous remarks can be sharpened by changing the temporal coordinate to $u = \ln |t|$. Note that $u$ now runs over all of $\R$, including zero. The effect of this change is simply $\Delta_P \mapsto \partial_u$. The Euler--Lagrange equations then take on a form in which the dynamics are a little clearer, 
\begin{align}\label{eqn:ELSigU}
&g_M^{ab}\nabla_b \nabla_a \Psi^i  +(1 - g_M^{ab}A_bA_a)\nabla_{u}\nabla_{u}  \Psi^i \\ \nonumber 
& +\Gamma^i_{jk}\left(g_M^{ab}\partial_b \Psi ^k \partial_a \Psi^j - 2 g_M^{ab}A_b \partial_a\Psi^k \partial_u \Psi^j + (1- g_M^{ab}A_a A_b)\partial_u \Psi^k \partial_u \Psi^j  \right)=0\,.
\end{align}
The temporal evolution is not (directly) governed by $t$, but rather the logarithmic time $u$. So long as solutions $ \Psi$ are sufficiently regular as discussed above, we have well-defined equations of motion for all $u \in \R$. \par 
A potentially physically interesting example is the following.  Consider the Schwarzschild black hole $(\mathcal{M} = \R^2 \times S^2, \rmd s^2)$, where we will employ Eddington--Finkelstein coordinates to write the spacetime interval as
$$\rmd s^2 = - \left(1 - \frac{2 G M_{BH}}{r} \right)\rmd v^2 + 2\, \rmd v \rmd r + r^2 \rmd \Omega^2\,,$$
where $\rmd \Omega^2$ is the round metric on the sphere $S^2$. Here $M_{BH}$ is the mass of the black hole, and $G$ is Newton's constant.  The event horizon is defined as $\mathcal{H} := \{ p\in \mathcal{M} ~~|~~ r(p) = 2GM_{BH}\} = S^2 \times \R$, which is a trivial line bundle. The induced degenerate metric on $\mathcal{H}$ is $g = 4G^2M_{BH}^2 g_{S^2}$, where $g_{S^2}$  is the round metric on $S^2$; note that the signature is $(1,1,0)$. Clearly, $\ker(g) = \Span(\partial_v)$. The vector field $\partial_v$ is a Killing vector field. \par 
The principal  $\R^\times$-bundle is given by removing the zero section, and the degenerate metric is $g$ now restricted to $v \neq 0$ To remain consistent with earlier notation, we will use $t$ as the fibre coordinate.  Clearly, the Euler vector field $\Delta =  t \partial_t$ spans $\ker(g)$; there is no dependence of $t$ in the degenerate metric $g$. Thus, we have a Carrollian $\R^\times$-bundle whose Euler vector field is Killing.  Setting the target space $N = \R$ with $k =1$, and noting that $P$ is trivial, we can select the trivial connection, so $A =0$.  The Lorentzian metric and the inverse are 
\begin{align*}
G = \begin{pmatrix}
(2G M_{BH})^2 \, g_{S^2}  & 0 \\
0  & - t^{-2}
\end{pmatrix}
\,,
& &
G^{-1} = \begin{pmatrix}
(2G M_{BH})^{-2} \, g^{-1}_{S^2}  & 0 \\
0  & - t^2
\end{pmatrix}  \,.
\end{align*}
The Euler-Lagrange equation \eqref{eqn:ELSigU} becomes
\begin{equation}\label{eqn:KGonBH}
\frac{1}{4 \kappa^2}\,\partial^2_u \Psi -  \Delta_{S^2} \Psi =0\,,
\end{equation}
where $\kappa = \frac{1}{4 GM_{BH}}$ is the surface gravity of the black hole, and $\Delta_{S^2}$ is the Laplace--Beltrami operator on $S^2$. Thus, the propagation speed with respect to the logarithmic time $u = \ln(|t|)$ is $\mathrm{v}= 2 \kappa$. Note, that we have non-trivial dynamics without a gauge field - something that is already novel in the Carrollian setting. The Hawking temperature (in natural units) is $T_H = \frac{\kappa}{2 \pi}$,  and so, 
$$\mathrm{v}= 2 \kappa = 4 \pi \, T_H\,.$$
General solutions to \eqref{eqn:KGonBH} are (using standard angular coordinates on $S^2$) 
$$\Psi(\theta, \varphi, u) = \sum_{l=0}^\infty \sum_{m = - l}^{l} \left( A_{lm} \cos(\omega_l \, u) + B_{lm} \sin(\omega_l \, u)\right)\, Y_{lm}(\theta, \varphi)\,,$$
where $\omega_l =  2 \kappa \sqrt{l(l+1)}$ and $Y_{lm}(\theta, \varphi)$ are the standard spherical harmonics on $S^2$. We make the following observations.
\begin{description}[itemsep=1em]
\item [Micro Black Holes] $M_{BH} \sim M_{Pl} \sim  10^{- 8}\, \text{--} \, 10^{-3}\,$Kg, $\kappa \gg 1$ (in natural units), `hot'; quantum gravity dominates. Temporal derivatives are suppressed compared to the spatial derivatives. The field $\Psi$ exhibits ultrarapid oscillations in $u$, and $\mathsf{v} \rightarrow \infty$ in the (formal) limit of complete evaporation.
\item [Macro Black Holes] $M_{BH}\sim 10^{9}\, \text{--}\, 10^{13}\,$Kg, $\kappa \sim \mathcal{O}(1)$ (in natural units), `warm'; threshold of where semiclassical gravity is valid. Neither temporal nor spatial derivatives dominated. The field $\Psi$ exhibits  balanced oscillatory behaviour, and $\mathsf{v}\sim \mathcal{O}(1)$.
\item [Astrophysical Black Holes] $M_{BH} \geq 10^{30}\,$Kg,  $\kappa \ll 1$ (in natural units), `cold'; classical general relativity dominates. Spatial derivatives are suppressed compared to the temporal derivatives. The field $\Psi$ becomes effectively ``frozen'' and  approaches a harmonic profile on $S^2$, with  $\mathsf{v} \rightarrow 0$ in the infinite-mass limit. 
\end{description}
%
%
\section{Concluding Remarks}
In this note, we have constructed  Carrollian sigma models using Carrollian $\R^\times$-bundles  rather than as a limit of a relativistic theory. An interesting and distinguishing feature of this model is the explicit presence of both spatial derivatives and temporal derivatives in the Lagrangian, ensuring that the dynamics are neither ultralocal (``electric'') nor solely dependent on spatial gradients (``magnetic''). This yields  a  dynamically rich class of Carrollian field theories that go beyond the standard theories constructed via extremal limits. While the physical relevance of such Carrollian theories remains unclear, it is intriguing to note that similar theories might arise in analogue gravity models. In particular, the fields may correspond to physical observables such as sound or phase perturbations in condensed matter systems.
%
%

%

\begin{thebibliography}{10}
\begin{small}

\bibitem{Bruce:2025}
Bruce, A.J.,
Carrollian $\R^\times$-bundles: Connections and Beyond, \href{https://doi.org/10.48550/arXiv.2505.21332}{ arXiv:2505.21332v2 [math.DG]}.

\bibitem{Ciambelli:2024}
Ciambelli, L.,
Dynamics of Carrollian scalar fields,
\href{https://doi.org/10.1088/1361-6382/ad5bb5}{\emph{Class. Quantum Grav.}}, \textbf{41}, No. 16, Article ID 165011, 17 p. (2024).  
 
\bibitem{Correa:2024}
Correa, F., Hernández, A. \& Oliva, J.,
The Carrollian limit of ModMax electrodynamics,
\href{https://doi.org/10.1007/JHEP12(2024)008}{\emph{J. High Energy Phys.}}, \textbf{12}, Paper No. 8, 13 p. (2024). 

\bibitem{Cotler:2024}
Cotler, J., Jensen, K., Prohazka, S., Raz, A., Riegler, M., \& Salzer, J.,
Quantizing Carrollian field theories,
\href{https://doi.org/10.1007/JHEP10(2024)049}{\emph{J. High Energy Phys.}},  \textbf{10}, Paper No. 49, 45 p. (2024). 


\bibitem{Duval:2014a}
Duval, C., Gibbons, G.W. \& Horvathy, P.A.,
Conformal Carroll groups,
\href{https://doi.org/10.1088/1751-8113/47/33/335204}{\emph{J. Phys. A, Math. Theor.}} \textbf{47}, No. 33, Article ID 335204, 23 p. (2014). 

\bibitem{Duval:2014b}
Duval, C., Gibbons, G.W. \& Horvathy, P.A.,
Conformal Carroll groups and BMS symmetry,
\href{https://doi.org/10.1088/0264-9381/31/9/092001}{\emph{Class. Quantum Grav.}} \textbf{31}, 092001 (2014). 

\bibitem{Duval:2014}
Duval,C., Gibbons, G.W.,  Horvathy, P.A. \&  Zhang, P.M., Carroll versus Newton and Galilei: Two Dual Non-Einsteinian Concepts of Time, \href{https://doi.org/10.1088/0264-9381/31/8/085016}{\emph{Class. Quantum Grav.}} \textbf{31}, 085016 (2014).

\bibitem{Ecker:2024}
Ecker, F.,  Grumiller, D., Henneaux, M., \& Salgado-Rebolledo, P.,
Carroll-invariant propagating fields, \href{https://doi.org/10.1103/PhysRevD.110.L041901}{\emph{Phys. Rev. D}} \textbf{110}, L041901 (2024).

\bibitem{Henneaux:1979}
Henneaux, M.,
Zero Hamiltonian signature spacetimes,
\emph{Bull. Soc. Math. Belg., Sér. A} \textbf{31}, 47--63 (1979). 



\bibitem{Lévy-Leblond:1965}
Lévy-Leblond, J.M.,
Une nouvelle limite non-relativistic du groupe de Poincaré,
\href{https://www.numdam.org/item/AIHPA_1965__3_1_1_0/}{\emph{Ann. Inst. Henri Poincaré, Nouv. Sér., Sect. A}} \textbf{3}, 1--12 (1965). 



\bibitem{SenGupta:1966}
Sen Gupta, N.D., On an analogue of the Galilei group, \href{https://doi.org/10.1007/BF02740871}{\emph{Nuovo Cimento A}} \textbf{44}, 512--517 (1966).

\bibitem{Sharma:2025}
Sharma, A., Studies on Carrollian Quantum Field Theories, \href{https://doi.org/10.48550/arXiv.2502.00487}{arXiv:2502.00487} (2025).
\end{small}
\end{thebibliography}
\end{document}